# Deep Learning-Assisted Fourier Analysis for High-Efficiency Structural Design: A Case Study on Three-Dimensional Photonic Crystals Enumeration

*Congcong Cui,[1] Guangfeng Wei,[1] Matthias Saba,[2,3],\* Yuanyuan Cao[4],\* and Lu Han[1],\**

[1] Tongji University, School of Chemical Science and Engineering, 1239 Siping Road, Shanghai, China

[2] University of Fribourg, Adolphe Merkle Institute, Chemin des Verdiers 4, Fribourg, Switzerland

[3] University of Fribourg, NCCR Bio-inspired Materials, Chemin des Verdiers 4, Fribourg, Switzerland

[4] School of Materials Science and Engineering, East China University of Science and Technology, Shanghai, China

E-mail: luhan@tongji.edu.cn; matthias.saba@unifr.ch; yuanyuancao@ecust.edu.cn

**Abstract:** The geometric design of structures with optimized physical and chemical properties is one of the core topics in materials science. However, designing new functional materials is challenging due to the vast number of existing and the possible unknown structures to be enumerated and difficulties in mining the underlying correlations between structures and their properties. Here, we propose a universal method for periodic structural design and property optimization. The key in our approach is a deep-learning assisted inverse Fourier transform, which enables the creation of arbitrary geometries within crystallographic space groups. It effectively explores extensive parameter spaces to identify ideal structures with desired properties. Taking the research of three-dimensional (3D) photonic structures as a case study, this method is capable of modelling numerous structures and identifying their photonic bandgaps in just a few hours. We confirmed the established knowledge that the widest photonic bandgaps exist in network morphologies, among which the single diamond (*dia* net) reigns supreme. Additionally, this method identified a rarely-known *lcs* topology with excellent photonic properties, highlighting the infinitely extensible application boundaries of our approach. This work demonstrates the high efficiency and effectiveness of the Fourier-based method, advancing material design and providing insights for next-generation functional materials.

# 1. Introduction

Structural design is the central topic of materials research, playing a pivotal role in determining material functionality and applications.[1,2] To date, numerous three-dimensional (3D) periodic structures have been discovered or engineered in both natural and artificial systems, exhibiting a wide range of functional properties (e.g., photonic, electric, magnetic, mechanical, energetic, etc.), which underscores the fundamental relationship between these diverse periodic structural features and material characteristics.[3,4] However, establishing efficient connections between all structural possibilities (no matter the existed or unknown ones) and their properties remains a significant challenge. The key to this challenge lies in how to enumerate all possible structural types, link their corresponding properties in a simple yet standardized form, and convert them into formats that can be easily integrated into databases for further description, analysis, and prediction.

While structural design varies with specific application requirements, the main design methods can be categorized into several key approaches: i) Database-driven methods draw inspiration from biological[5] or crystallographic databases such as Reticular Chemistry Structure Resource;[6] ii) Basic unit manipulation involves strategically positioning building blocks at specific coordinates or interconnecting them;[7,8] iii) Heterogeneous designs embed variations in composition, reinforcement patterns, or phases within the same structural framework;[9,10] iv) Multiscale designs integrate features across different length scales or graded transitions;[11,12] v) Topology optimization iteratively refines geometric layouts to enhance performance while minimizing weight or maximizing efficiency,[13,14] etc. However, current structural designs are largely anchored to known configurations or initial guesses, thereby restricting the possibilities of creating new structures beyond prior experience and established knowledge. Notably, even for periodic structures, with their 230 crystallographic space groups in 3D, there remains an immense variety of possible crystal structures that could be explored. For a given functional material type (for example, the 3D photonic crystals), a majority of 3D symmetrical architectures remain unexplored, representing a largely untapped design space for functional materials. Although pixelated modeling offers maximum structural freedom,[15,16] the common method to generate 3D architectures in reality necessitates intricate descriptions for individual pixels, leading to an exponential increase in parameters and computational cost as complexity increases.[17] Yet, developing a design methodology that can generate all structural possibilities within a practically relevant physical design space remains challenging.

We, here, propose such a method that can additionally predict the physical properties of each structure through an ultrafast computational scheme. The key in our structural design method is based on Fourier analysis to address the intractable structure enumeration problem. It is known that the Fourier transform decomposes a periodic function into a series of sine and cosine components. Conversely, the inverse Fourier transform reconstructs the original periodic signal from these frequency components. Fourier analysis has been extensively employed in crystallography to determine unknown crystal structures represented as a periodic function of their spatial electron density distributions.[18,19] By measuring the Fourier coefficients corresponding to the Bragg lattice planes (also known as crystal structure factors) through X-ray diffraction experiments and phase fitting,[20] the crystal structures in real space can be reconstructed. In particular, these experiments allow calculating the spatial electron density distribution $\rho_{(r)}$ of each fractional coordinate $r \in [0,1)^3$ in the unit cell to construct the level set according to the equation[21,22]

$$\rho_{(r)} = \frac{1}{V_c} \sum_{k \in M} F_k \exp\left(-2\pi i \mathbf{k} \cdot \mathbf{r} + \alpha_k\right) \tag{1}$$

where $V_c$ is the unit cell volume, $M$ is a subset of $\mathbb{Z}^3$ and represents a set of crystal faces with corresponding reflection conditions in the lattice of the given space group. $F_k$ and $\alpha_k$ are the amplitude and phase of the crystal structure factors corresponding to the $k$ Miller index Bragg reflection.

This basic mathematical statement can construct a library containing all diverse and topologically complex 3D crystal structures associated with the 230 space group symmetries, no matter for the conventional crystals formed by discrete atoms/molecules or for the unconventional meso/macro-structural assembled structures enclosed with infinite continuous surfaces or metastructures. This unified description allows the systematic exploration of diverse geometry designs for materials composed of two (or more) components with sharp yet smooth (analytical) interfaces, which is particularly well-suited for addressing almost all engineering problems of interest. Furthermore, this method facilitates a deeper physical understanding of corresponding material functionalities through comprehensive databases and simple structural parameters—the crystal structure factor amplitudes and phases. Such a complete characterization from a mathematical perspective is vital for the accurate calculation of corresponding materials functionalities.

However, the sheer scale of large databases still imposes an enormous computational workload. In the conventional material explorations, even slight structural variations require additional experimental measurements or simulations. Therefore, elucidating the relationships between the structure and activities becomes extremely time-consuming and labor-intensive, far exceeding the capabilities achievable through manual approaches. To address this challenge, we take the power of deep learning (DL)[23-25] in handling the big data of the structural design parameters and the corresponding performance parameters. Furthermore, we incorporate Non-dominated Sorting Genetic Algorithm II (NSGA-II), an intelligent 'multi-goal optimizer' that balances competing design objectives, into our framework for inverse structural design.[26] Using this approach, we can efficiently connect the structural database obtained through Fourier analysis with their numerically simulated performance characteristics. The synergy between Fourier analysis and machine learning not only accelerates the discovery of novel materials but also empowers inverse design, where desired material properties guide the generation of optimal structural configurations.

## 2. Results

### 2.1. Structural modeling by inverse Fourier transform

A wide range of materials in nature and artificial systems, including photonic, mechanical, phononic, and electrical materials, have close relationship with 3D periodic structures. As a representative case study, we apply this method to explore the 3D photonic crystals to illustrate the advantages of our approach. Photonic crystals are periodic, binary refractive index modulations that confine and control the propagation of light, which have significant implications for optical applications such as inhibiting spontaneous emission, guiding and bending of light, optical computers and information devices, etc.[27-30] The photonic bandgap in photonic crystals is highly dependent on their structural design. Specifically, identifying and optimizing 3D photonic crystal structures with omnidirectional complete photonic bandgaps (CPB) remains critical for overcoming the current performance limitations of optical devices.[31,32] By using this method, any photonic crystal structure can be represented by its Fourier coefficients (analogous to crystal structure factors), allowing for the systematic exploration of structural complexity and its impact on optical properties. Different from the traditional topological optimization of photonic crystals,[13,14,33] which often struggles to identify the global minimum without an informed initial guess due to the energy landscape of the cost function is generally highly non-convex, this method generates a database of all possible structural types, from which we could predict the band structures or deduce the structure from the desired properties.

Figure 1A shows the schematic diagram of a general Fourier-DL workflow in exploring the structure of photonic crystals. The symmetry elements defined by the space group under consideration dictate a specific set of systematic extinction conditions of Bragg reflections[34] during the inverse Fourier transform process. Subsequently, the 3D spatial electron density distribution is calculated using randomly generated crystal structure factor amplitudes and phases, where these parameters are uniformly distributed over their respective ranges. The final structure is determined by the normalized electron density $\rho_{norm\ (r)}$, rescaled to the range of 0–100%, along with the equi-electron density level threshold $t$,

$$\rho_{norm\ (r)} = \frac{\rho_{(r)} - \rho_{min}}{\rho_{max} - \rho_{min}} > t \qquad (2)$$

where $t$ represents the isosurface threshold for determining the structure boundary, which adjusts the fill fraction of the two dielectric materials. Thus, the allowed amplitudes and phases of the Fourier series and the threshold $t$ serve as structural parameters whose values can be adjusted to model arbitrary periodic photonic structures in specified space group symmetries (Figure 1B). The corresponding photonic band structures are obtained through electromagnetic simulation employing MIT Photonic-Bands (MPB)[35] with a corresponding spatial resolution of 16×16×16 for the initial search and 32×32×32 for the computation of specific structures. Subsequently, the deep neural network is trained using the dataset constructed of 100 000 sets of structure-performance data of selected space groups to efficiently predict the photonic band structure from the structural parameters (Figure 1C). Then, a deep neural network acts as a proxy model for the electromagnetic simulation process, combined with a genetic algorithm to achieve an efficient search of the structural design space and obtain the photonic crystal structural parameters with specified band characteristics (Figure 1D).

Generally, the formation of a CPB is the result of coherent scattering of electromagnetic waves by the photonic crystal.[29,30] The scattering process should, however, not be understood in the form of a Born series, which does generally not converge for practical refractive index contrasts. Instead, a classification with respect to the translation symmetry of the crystal using Bloch's theorem can explain the occurrence of a bandgap through the absence of available photonic states in the material.[27] A large CPB requires an absence of states in all directions of propagation, which can be addressed by structures with the most spherical Brillouin zone.[36] Additionally, connected low-coordination-number networks of uniform valency possessing substantial local self-uniformity are considered conducive to the formation of large CPBs.[37,38]

To date, the face-centered cubic single diamond (SD, *dia* net, space group $Fd\bar{3}m$, No. 227) structure with tetravalent vertices is considered the "holy grail" of all photonic structures due to the most isotropic and highly geometrical identical framework, exhibiting the widest photonic bandgap and the lowest dielectric contrast requirement.[31,37,39] The body-centered cubic, chiral single gyroid (SG, *srs* net, space group $I4_132$, No. 214) structure with trivalent nodes, which also exhibits perfect local self-uniformity, also forms a large CPB comparable to but smaller than the diamond.[37,40] These two structures have also been found to be the source of structural colors in various biomineralized skeletons and scaffolds, in which the dielectric contrast is insufficient to open a CPB.[41-44] In addition, the single primitive (SP, *pcu* net, space group $Pm\bar{3}m$, No. 221) network also produces a relatively smaller CPB in simulations,[40,45] which can be attributed to the fact that it consists of hexavalent nodes with less local self-uniformity.[38] Considering the excellent photonic properties of these structures, we apply our method to the corresponding space groups.

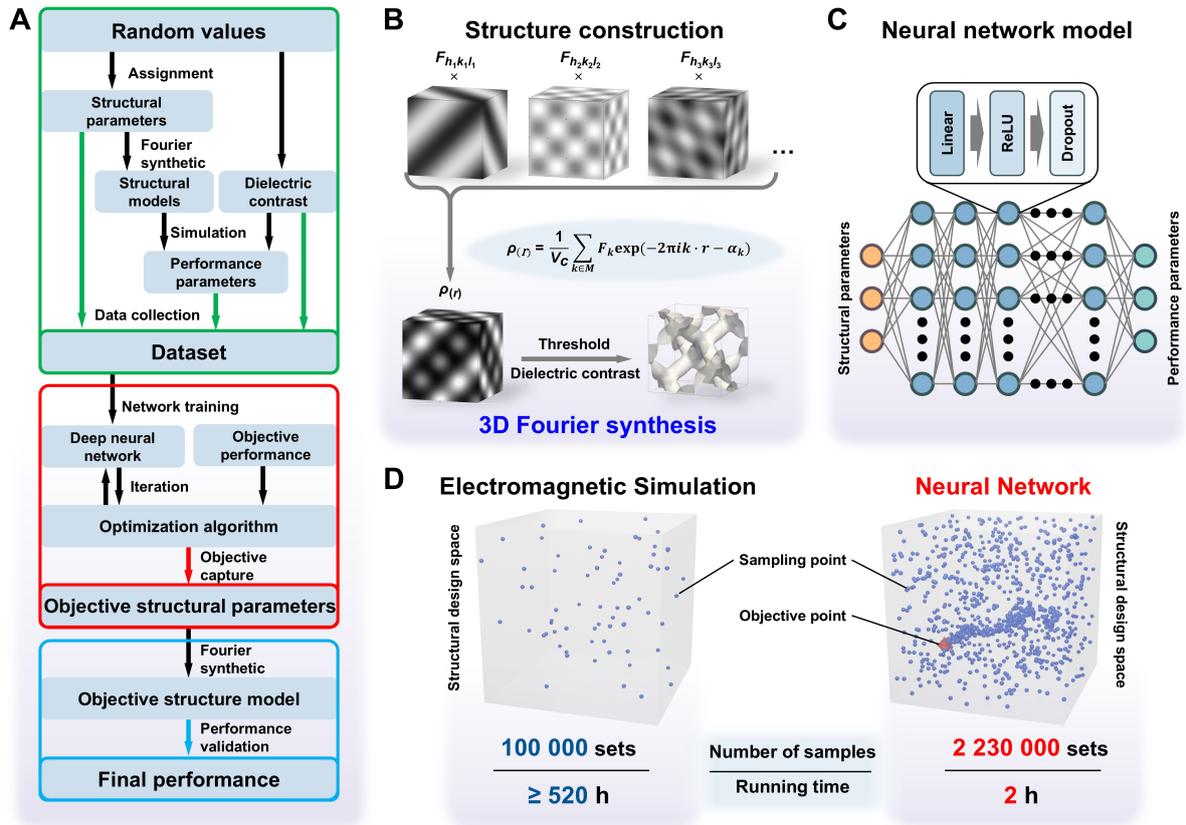

**Figure 1.** Schematic workflow of the construction model by Fourier analysis and the structural design using a deep neural network. (A) Schematic diagram of the workflow. A random sampling of the structural design space was achieved by inverse Fourier transform with random parameters in Equation 1 combined with the dielectric contrast to calculate the performance parameters by electromagnetic simulation. The structural parameters, dielectric contrast and performance parameters were collected to construct a dataset (green box) to train the deep neural network. The deep neural network was combined with genetic optimization algorithm to capture the objective structural parameters (red box) through an iterative process with a predefined performance objective. The objective structure was constructed again by inverse Fourier transform and its performance was verified by electromagnetic simulation (blue box). (B) The construction of photonic structures by inverse Fourier transform. (C) The neural

network model architecture using a multilayer perceptron. The structural parameters and performance parameters (photonic band structure) were used as input and output data of the neural network, respectively. Bandgap performance can be obtained from the band structure through simple post-processing. The Rectified Linear Unit (ReLU) function was used as a nonlinear activation and a dropout layer was added to prevent overfitting after each hidden layer. (D) Comparison of the efficiency of electromagnetic simulation versus neural network for the exploration of the structural design space. The sampling of 100 000 sets of samples by electromagnetic simulation for the structural design space took more than 520 h. The neural network took only 2 h to explore the structural design space for 2.23 million sets of samples during the structural design process. The efficiency of sampling the structural design space is, therefore, improved by more than three orders of magnitude.

We also consider the space groups of their correlated double-network morphologies, namely, of the self-dual double diamond (DD, *dia-c* net, space group $Pn\overline{3}m$, No. 224), the achiral double gyroid (DG, *srs-c* net, space group $Ia\overline{3}d$, No. 230) and self-dual double primitive (DP, *pcu-c* net, space group $Im\overline{3}m$, No. 229) network. The multi-network structures are capable of demonstrating fascinating phenomena, such as 3D Weyl points and large circular polarization stop bands.[46,47] However, it has been proposed that double-networks are unlikely to produce CPBs.[40] Even so, exploring the photonic structure within these symmetries remains appealing. We, therefore, concentrate our explorations in this paper using these six space group symmetries. We build the structural models by inverse Fourier transform using structure factors $F_{hkl}$, applying the calculation rules in *International Tables for X-Ray Crystallography*.[19] Five representative low-index Bragg reflections (abbreviated as reflections) were chosen for effective conceptual validation of our workflow. Introducing more reflections as parameters can generate topologically more complex structures of higher genus. Additionally, high-index reflections contribute more to geometrical detail, while low-index diffraction patterns are more decisive in determining the overall structural features. Herein, highly symmetric cubic space groups were selected and the scattering length scale is on the order of the unit cell. Under the assumption that the bandgap does not critically change with geometrical perturbation, we focused our calculation using five low index representative reflections. This allows us to model most structural possibilities within the space group symmetry while achieving high computational efficiency. Each unique *hkl* reflection corresponds to a group of symmetrically equivalent {*hkl*} Fourier coefficients, which possess identical amplitudes but their phases are related through symmetry operations. These are listed in *International Tables for X-Ray Crystallography* for all space groups. Prior to substitution into the calculation equations, each unique reflection was expanded to encompass all symmetry-equivalent reflections. Notably, for centrosymmetric structures, the structure factor phase can only be 0 or $\pi$, which is reflected in the calculation through the sign of the structure factor amplitude, where the positive numbers correspond to a phase of 0 and the negative value correspond to a phase of $\pi$. Since the space group $I4_132$ (214) is non-centrosymmetric, its phase combinations will result in an excessively large structural design space. Consequently, the phases were referenced to the SG structure and confined to 0, $\pi/2$, $\pi$, and $3\pi/2$. This setting somehow restricts the structural types within this space group but ensures the efficient generation of structures with complete bandgap. The selection and reflections for the space groups and the calculation equations are listed in Table S1-S7. The crystal structures modeled with different structural parameters in each space group are illustrated in Figure S1-S6.

The structural model can be effectively tuned by adjusting the values of the structural parameters (Figure 2). Figure 2A and Animation S1 illustrate how varying the threshold value $t$ in the space group $Fd\bar{3}m$ (227) with only the 111 reflection present ($F_{111} = 1.0$, $F_{220} = 0.0$, $F_{311} = 0.0$, $F_{222} = 0.0$, $F_{400} = 0.0$, which are the five lowest reflexes in space group $Fd\bar{3}m$) affects structure formation. As $t$ is the normalized threshold value in the range of [0, 1) as defined in Equation 2, the material interface resembles the diamond triply periodic minimal surface (same symmetry $Pn\bar{3}m$ and topology) for $t = 0.5$. For $t = 0.3$ and 0.7, the structure represents a SD network topology with complementary volume fill fraction. By continuously increasing $t$ to 0.9, it transforms into simple structural units (rounded tetrahedra), orderly distributed at diamond lattice coordinate sites. As shown in Figure 2B and Animation S2-S5, by independently adjusting other coefficients based on the fixed structural parameters of $F_{111} = 1.0$ and $t = 0.5$, we were able to generate a variety of complex structural models. These include many complex geometries that cannot be constructed using simple structural units (e.g., spheres, cylinders, etc.). Such structural variations lead to corresponding changes in their photonic properties. The band structures of these models calculated under a dielectric contrast of 13.00 are also shown in the animations. The structural diversity, design flexibility and complexity of the Fourier analysis-based modeling approach are illustrated in Figure 2.

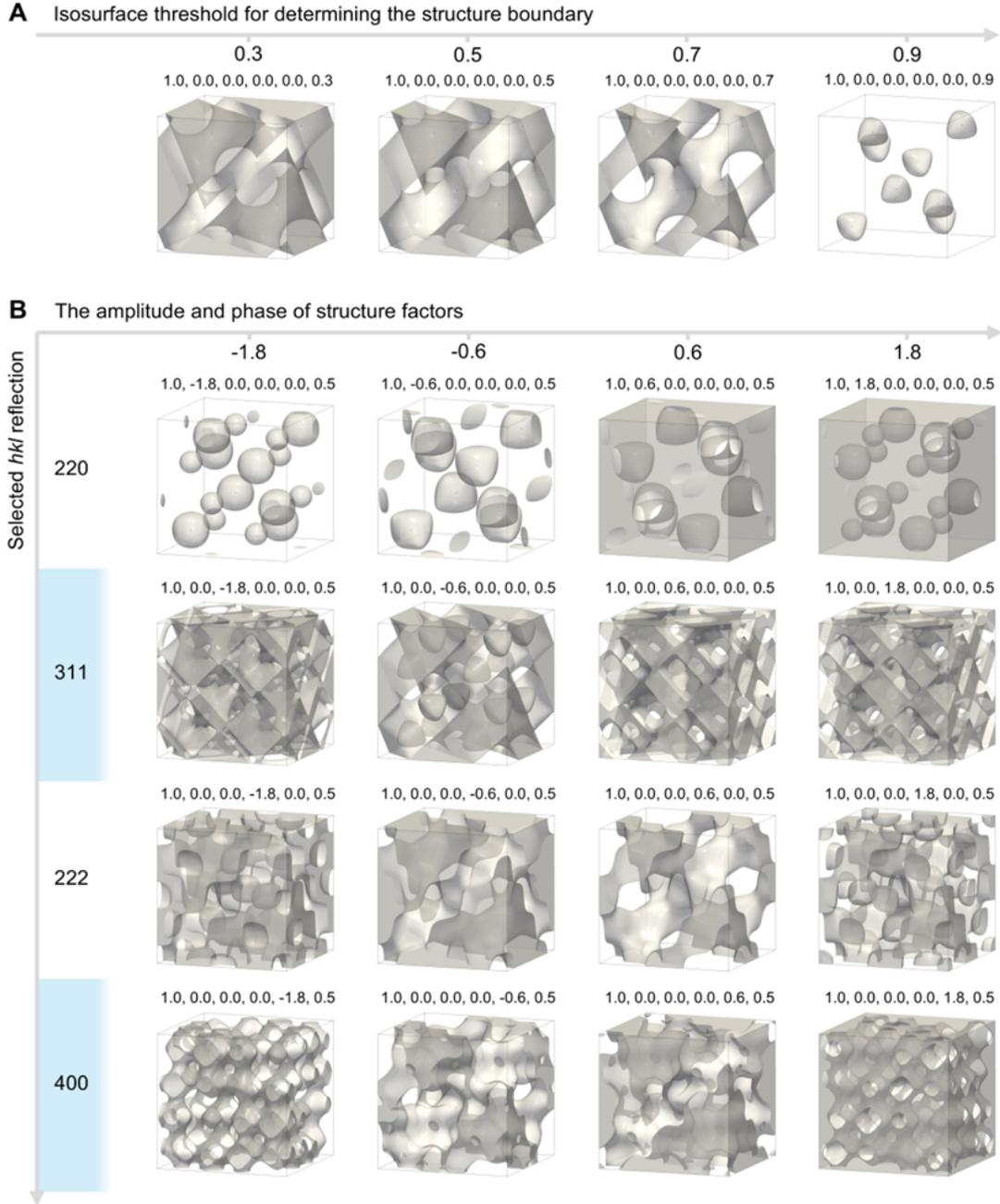

**Figure 2.** Structural models under different model parameters. (A) Structural models generated with $F_{111} = 1.0$, $F_{220} = 0.0$, $F_{311} = 0.0$, $F_{222} = 0.0$ and $F_{400} = 0.0$ using different isosurface thresholds. (B) Structural models obtained with fixed structural parameters of $F_{111} = 1.0$ and $t = 0.5$ by adjusting other structure factors. The value of of $F_{111}$, $F_{220}$, $F_{311}$, $F_{222}$, $F_{400}$ and threshold $t$ are shown on the top of each model.

## 2.2. Exploration of photonic bandgap performance space

The structural parameter range was determined by comparing the proportion of structures with CPB (>1%) under different structural parameter settings. Taking the space group $Fd\bar{3}m$ (227) as an example, 1 000 candidate structures were generated by random structural parameters of five low-index reflections. The corresponding band structures (performance space) were calculated using MPB with the dielectric contrast of 13.00 for the filling material. As shown in Table S8, the probability of photonic structures with CPB varies with the data range of the five reflections. The highest probability occurred when the amplitude of the first reflection was set to 1 and the other four reflections were in the range of [-1, 1), where ~15% of the 1 000 structures exhibited a CPB. In this scheme, the first reflection provides the largest contribution to the structure, and other reflections would modulate the structure on this basis. Therefore, this setting was used in subsequent calculations to ensure computational efficiency.

Within the parameter range of the above-mentioned design space, 10 000 models were randomly sampled for each of the six space groups. The corresponding band structures were calculated by MPB. CPB structures were found in all space groups (Figure 3 and Table S9). Particularly, the space group $Fd\bar{3}m$ possessed over 1 700 CPB structures and showed a maximum gap between frequencies (*f*) 0.5111-0.6958 *c/a* with a gap width of 30.60%. The $I4_132$ structures also showed excellent photonic properties with the widest gap width of 26.89% between *f* range 0.4436-0.5814 *c/a*. The CPB structures also existed in other space groups, in particular, nearly 1 000 CPB structures were identified in space group $Ia\bar{3}d$. Although the widest gap width (23.75%) was lower than that of the $Fd\bar{3}m$, the *f* range 0.7228-0.9176 *c/a* was significantly higher than that of $Fd\bar{3}m$. Its CPB occurs at a very high band index between band 12 and 13. In fact, the connectivity index based on crystal symmetry alone, only allows a bandgap to open above the 8th band in $Ia\bar{3}d$.[49] Additionally, the $Ia\bar{3}d$ structure is geometrically more complex, with a genus of 13 per primitive unit cell, compared to a genus of 3 for the gyroid and the diamond. The relevant scattering dimension for destructive interference is therefore smaller compared to the unit cell, further explaining the occurrence of the bandgap at higher frequencies. For the other three space groups, the number of structures capable of forming CPBs was less than 3% of the total structures, and the gap width obtained was below 18%. In comparison, the space group of the more common space group $Fm\bar{3}m$ (No. 225) hosted 4.05% CPB structures in all models with a maximum gap width of 15.84% (Figure S7).

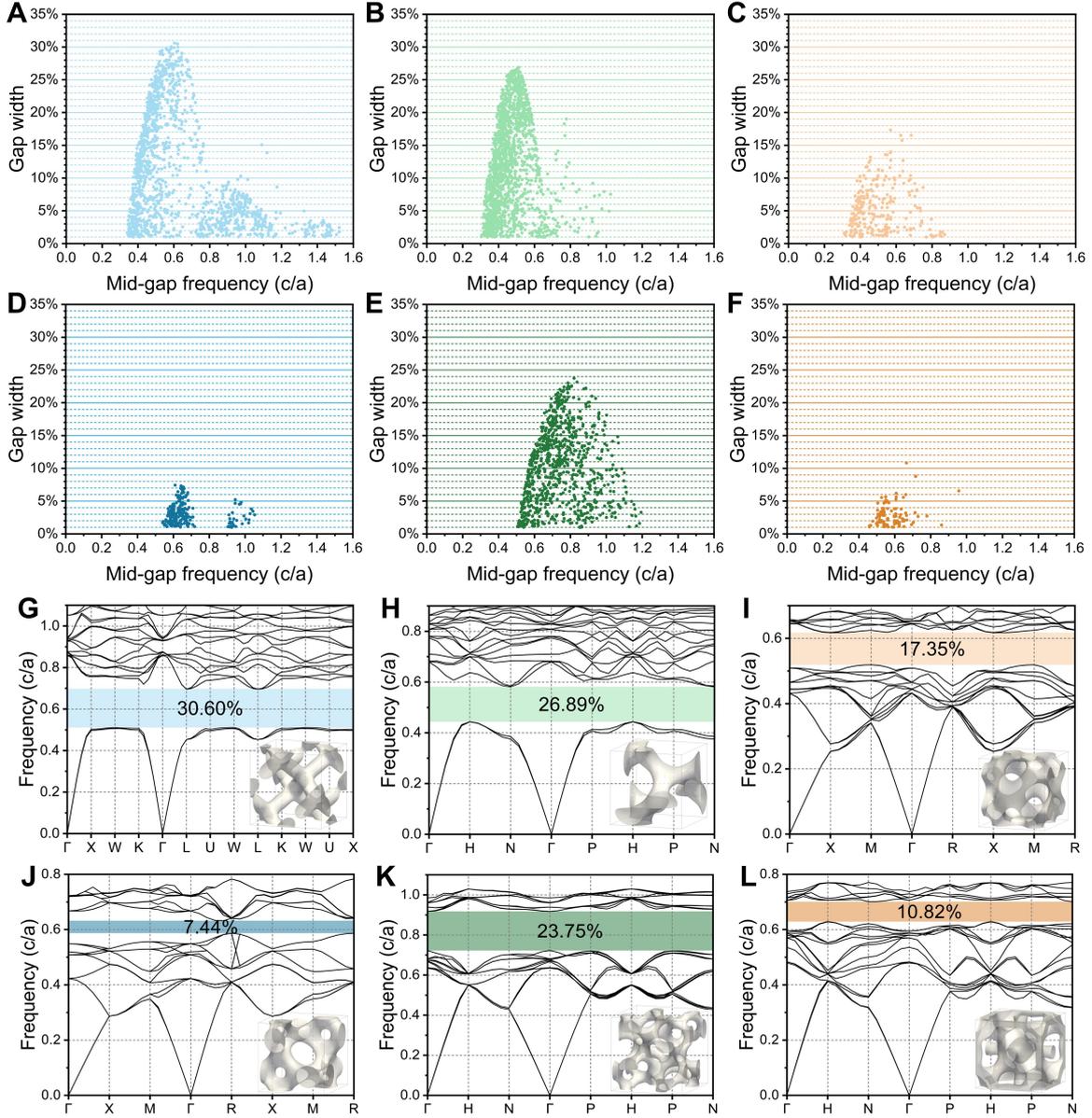

**Figure 3.** Photonic bandgap properties of different space groups. (A-F) Performance space for each space group of (A) $Fd\bar{3}m$, (B) $I4_132$, (C) $Pm\bar{3}m$, (D) $Pn\bar{3}m$, (E) $Ia\bar{3}d$, (F) $Im\bar{3}m$ with dielectric contrast of 13.00. (G-L) Structures with the largest gap width found in each space group and their band structures, (G) $Fd\bar{3}m$, (H) $I4_132$, (I) $Pm\bar{3}m$, (J) $Pn\bar{3}m$, (K) $Ia\bar{3}d$, (L) $Im\bar{3}m$. The *k*-path refers to Stefano Curtarolo's suggestion.[48]

## 2.3. Exploration of bandgap properties and design of structures with $Fd\bar{3}m$ symmetry

Considering the excellent photonic properties of the $Fd\bar{3}m$ symmetry, we further incorporated the dielectric contrast from 1.00 to 16.00 into the structural design parameters. 100 000 models with different parameter combinations were constructed, and the corresponding photonic properties were calculated. As shown in Figure S8, a minimum dielectric contrast of 3.87 was required to open a CPB, and the gap width increased with increasing the dielectric contrast. Among the 100 000 models, 8 392 samples yielded a CPB greater than 1%, with the mid-gap frequency ($f_{mg}$) in the range of 0.3091-1.2433 $c/a$ (Figure 4A). Among them, the structure possessing the widest bandgap (with detailed structural parameters presented in Figure S9) exhibits a 34.55% CPB in the $f$ range of 0.4735-0.6713 $c/a$ accompanied by a dielectric contrast of 15.35 and a gap width of 30.88% in the $f$ range of 0.5089-0.6948 $c/a$ with a dielectric contrast of 13.00. We further confirmed the bandgap of this structure using a high resolution of 32×32×32 and obtained consistent results (34.99% and 31.31%, respectively). It is worth noting that, both the structure with the largest gap width and the CPB structure with minimum dielectric contrast are related to the SD structure with *dia* topology, albeit with different structural parameters. Judging from the results of the 100 000 samples, the performance parameters shown in Figure 4A should be close to the upper limit of optimized performance.

A fully connected multilayer perceptron (MLP)[24,50] was trained using the dataset consisting of 100 000 samples described above to enable direct prediction of the corresponding band structures from the structural design parameters. The tested MLP hyperparameters with the corresponding losses and the loss for the validation dataset of the optimal MLP model are shown in Figure S10 and S11. The MLP successfully predicted 794 out of the 882 structures with CPB in the test dataset. An example of the predicted band structure is shown in Figure 4B (structural parameters are listed in Table S10), which is highly consistent with the electromagnetic simulation. Moreover, the predicted $f_{mg}$ and gap width concentrated around the ideal electromagnetic simulation results (Figure 4C and Figure S12).

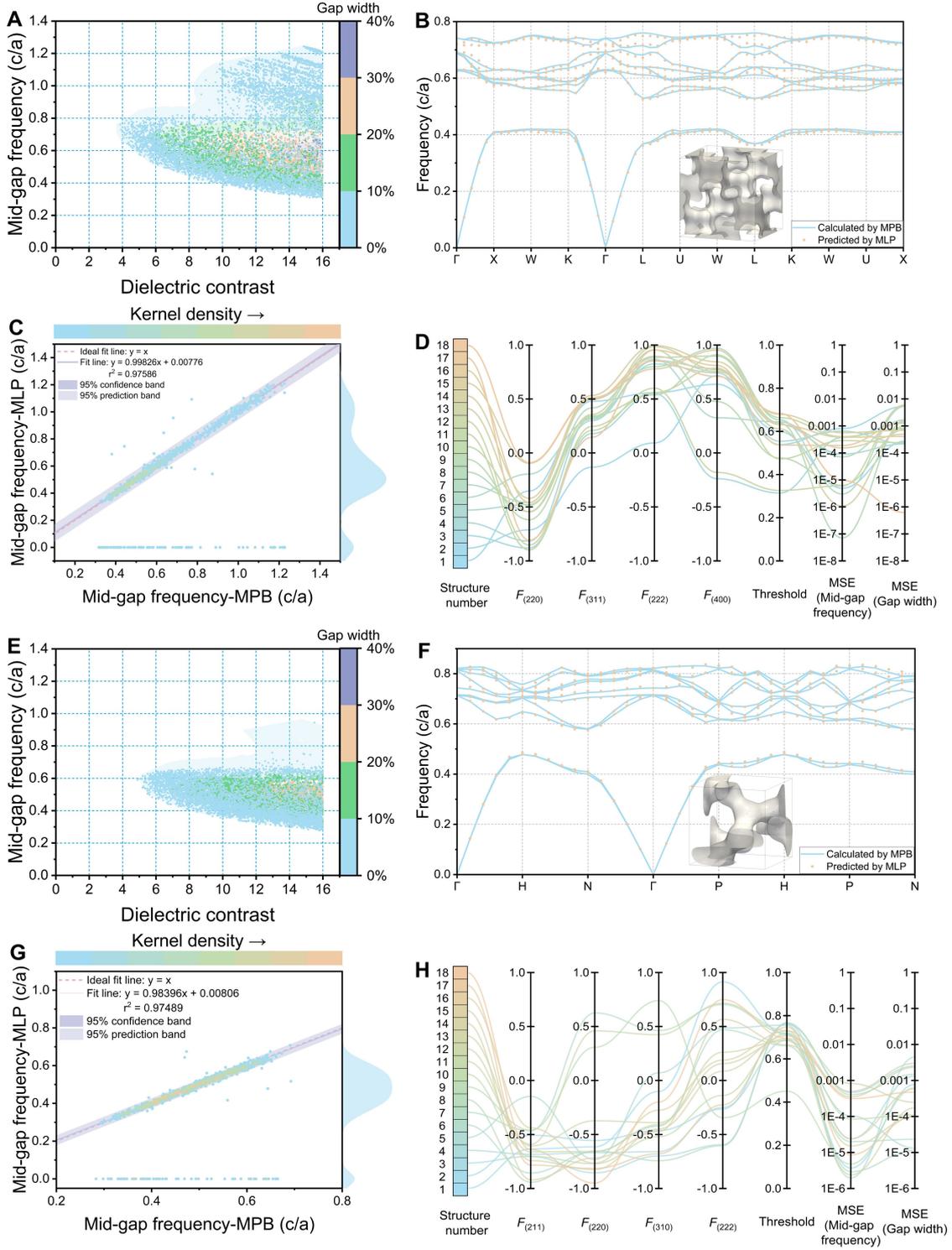

**Figure 4.** Prediction and inverse design of photonic crystal performance based on MLP of $Fd\bar{3}m$ and $I4_132$ symmetries. (A) Design space of photonic bandgap performance parameters in the dataset of $Fd\bar{3}m$. (B) Photonic band structure of one representative sample in the test dataset calculated by MPB (blue line) and predicted by MLP (orange dots) in the space group $Fd\bar{3}m$ with a dielectric contrast of 15.98. The inset shows the structural model. (C) Correlation between the $f_{mg}$ calculated by MPB and the predicted $f_{mg}$ by MLP in the test dataset of $Fd\bar{3}m$.

In this case, $f_{mg} = 0$ means that there is no gap in the band structure. Samples that both have bandgap are used to compute the fit line. (D) Structural parameters of the photonic crystals with $Fd\bar{3}m$ symmetry obtained by MLP inverse design and the corresponding photonic bandgap performance with the mean-square error (MSE) of the target performance. (E) Design space of photonic bandgap performance parameters in the dataset of $I4_132$. (F) Photonic band structure of a representative structure calculated by MPB (blue line) and predicted by MLP (orange dots) in the space group $I4_132$ with a dielectric contrast of 10.11. The structural model is shown as inset. (G) Correlation between the MPB calculated $f_{mg}$ and the MLP predicted $f_{mg}$ in the test dataset of $I4_132$. (H) Structural parameters of $I4_132$ obtained by MLP inverse design and the corresponding bandgap with the MSE of the target performance.

The trained MLP combined with the genetic algorithm is capable of searching and iteratively optimizing the geometric structural design parameters with the target bandgap performance. For inverse structural design, the $f_{mg}$ (0.6 $c/a$) and the gap width (20%) were chosen as two optimization objectives as an example. The dielectric contrast 13.00 of the material was used as a constraint. The structural design parameters (the structure factors shown in Table S1 and the threshold $t$) were iteratively optimized (Figure S13). 2.23 million sets of structural parameters were evaluated within 2 h using a desktop workstation containing two 3.2-GHz Intel(R) Xeon(R) 6146 CPUs and an NVIDIA Quadro P2000 GPU. The band structures of the resulting geometries were computed with MPB to verify the performance of the optimized geometries. Figure S14 shows the photonic bandgap performance for 18 sets of parameters obtained from the geometrical design space. As shown in Figure 4D, S15 and S16, the results predicted by the MLP were close to those calculated by the electromagnetic simulation, and their performances met the original design requirements well. Interestingly, the resulting structures were diverse, including not only network structures but also hybrid structures with disconnected geometrical domains coexisting with percolating networks.

### 2.4. Exploration of bandgap properties and design of structures with $I4_132$ symmetry

As shown in Figure 3B, the space group $I4_132$ also showed diverse possibilities of well performing photonic structures. 8 616 samples with CPB greater than 1% were found among 100 000 sets of structural parameters calculated by electromagnetic simulation. As shown in Figure 4E and S17, the minimum dielectric contrast of 4.89 was identified in space group $I4_132$ for a CPB to appear, significantly larger than the value of 3.87 found for the space group $Fd\bar{3}m$. The widest gap of 31.87% in space group $I4_132$ was generated from a structure characterized by 110 and 211 reflections with similar amplitude but opposite phase (specific structural parameters are shown in Figure S18) at a dielectric contrast of 15.85 with the $f$ range between 0.4027 and 0.5554 $c/a$, pointing to a SG-like structure with *srs* topology. The gap width of the structure was 26.78% in the $f$ range 0.4400-0.5760 $c/a$ for dielectric contrast of 13.00. These results were also consistent with the resolution of 32×32×32 (31.64% and 26.83%, respectively). Using the same workflow as for $Fd\bar{3}m$, a fully connected MLP was trained to map the structural parameters to the band structure. The validation dataset loss of the model training process is shown in Figure S19. The photonic band structures predicted by the MLP were in high agreement with the electromagnetic simulations (Figure 4F and the structural parameters are shown in Table S11). For the 877 sets of samples with CPBs in the test dataset, 780 of them were similarly predicted. As shown in Figure 4G and Figure S20, the predicted $f_{mg}$ and gap width mainly concentrated around the electromagnetic simulation results. Based on the trained MLP, the inverse design of photonic crystal structures can, therefore, be applied. Taking the $f_{mg}$ (0.5 $c/a$), gap width (15%), and dielectric contrast 13.00 as targets, the 18 sets of samples obtained through optimization well satisfied the original design requirements (Figure 4H, S21-S23).

## 2.5. Photonic structures with space group $Ia\bar{3}d$ as rivals of "holy grail" photonic crystal

Notably, the crystal structures with space group $Ia\bar{3}d$ also showed excellent photonic properties with high $f_{mg}$ (Figure 3E). Therefore, 100 000 sets of random structural parameters and photonic bandgap performance parameters were generated by electromagnetic simulation. Among them, 6 232 sets of samples showed CPB greater than 1% with the $f_{mg}$ between 0.4642-1.2357 $c/a$ (Figure 5A). Particularly, the $f_{mg}$ of the samples with gap widths greater than 20% ranged from 0.6106 to 0.9782 $c/a$. It is worth noting that a minimum dielectric contrast of 3.94 was required to ensure a CPB for $Ia\bar{3}d$ structure (Figure 5B, S24 and S25), which is very close to the value of the $Fd\bar{3}m$-based structures of 3.87. This structure consists of a connected network with nodes at (0.000, 0.250, 0.375) with $Ia\bar{3}d$ space group symmetry and the topology of an *lcs* network (Figure 5C). It can be described as an infinite tiling of space along four <111> directions by tiles composed of five hexagonal faces (Figure 5D).[51] Compared to the structures with $I4_132$ symmetry, this space group possesses much higher symmetry, the structure is achiral with complex connectivity and smooth saddle-shape surface. Notably, the structure with the maximum gap width was discovered with 321 and 400 reflections having a large amplitude with the opposite phase to the fundamental 211 reflection (the structural parameters are shown in Table S12). Under a computation resolution of 32×32×32 in MPB, this structure creates a gap width of 24.75% between $f$ range 0.7409-0.9501 $c/a$ with a dielectric contrast of 14.31 and a gap width of 23.04% in the $f$ range 0.7722-0.9732 $c/a$ with a dielectric contrast of 13.00. Using the same dielectric contrast of 15.35 as for the $Fd\bar{3}m$ structure, the gap width increased to 25.78% in the range of 0.7195-0.9325 $c/a$ (Figure 5E). Due to the relatively high complexity of this structure, the results calculated at higher resolution are more accurate. For comparison, the gap widths are 25.91%, 24.17% and 27.72% for dielectric contrasts of 14.31, 13.00 and 15.35, respectively, calculated with a mesh size of 16×16×16. Although the gap width is still smaller than that of $Fd\bar{3}m$, the gap frequency range is improved by 7.2% (Figure 5F).

There is an ostensible increase in $f_{mg}$ compared to the diamond-like morphology in the *lcs* structure, which suggests that it reduced the requirements for manufacturing precision in top-down fabrication. However, the *lcs* topology is structurally more complex within the unit cell, with vertices on the 24d Wyckhoff positions (compared to only 8a in the case of the diamond), and 48g mid-edge positions (compared to 16c), resulting in a genus of 13 (number of rings) per primitive unit cell (compared to a genus of only 3 for the diamond). The *lcs* and the *dia* net are both 4-coordinated and lead to tiling with 6-ringed facets. They only differ in the number of facets per tile (5 in case of the *lcs* and 4 in case of the self-dual *dia*), and the bond angles (100 and 132 degrees for the *lcs* and 110 degrees for the *dia*). Given the similarities of these two nets, the structural complexity is best covered by a frequency that is measured in units of $c/d$ instead of $c/a$, where $d$ is the edge length of the network. In these units, the $f_{mg}$ of the *lcs* bandgap at refractive dielectric contrast 13.00 is 0.2672, and that for the *dia* is 0.2591. Even though the advantage in fabrication complexity is small, the bandgap opens at a similar refractive dielectric contrast for the *dia* and the *lcs* net, warranting further investigation of the latter. Meanwhile, it seems at least possible that more complex 3 or 4-coordinated geometries can rival or even beat the diamond as the reigning photonic bandgap champion structure.

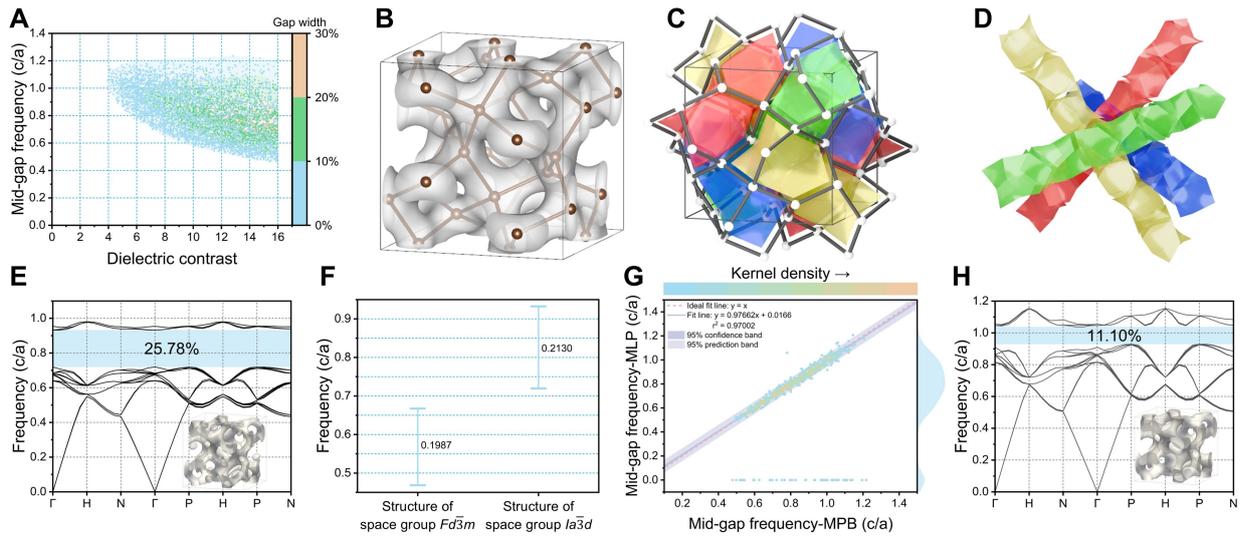

**Figure 5.** Prediction and inverse design of photonic crystal performance based on MLP of $Ia\bar{3}d$ symmetry. (A) Design space of photonic bandgap performance parameters in the dataset. (B) CPB photonic structure with the minimum requirement of dielectric contrast (3.94) of $Ia\bar{3}d$. (C) Tessellation model of the structure in (B). (D) Fragments of the *lcs* tiling. (E) The photonic band structure with the maximum gap width in the dataset (dielectric contrast of 15.35) and the corresponding geometry. (F) Photonic gap frequency range of the structures in $Fd\bar{3}m$ (Fig. S9A) and $Ia\bar{3}d$ (E), both with dielectric contrast of 15.35. (G) Correlation between the mid-gap frequency calculated by MPB and those predicted by MLP in the test dataset. (H) Photonic band structure and geometrical structure with maximum gap width obtained by inverse design with dielectric contrast of 6.25.

The MLP was trained based on the 100 000 sets of geometrical structural parameters and photonic bandgap data. The validation dataset loss of the model training process is shown in Figure S26. As shown in Figure 5G and Figure S27, 651 sets of samples of interest showed CPB in the test dataset, and the $f_{mg}$ of 597 sets of predicted structures were highly consistent with the electromagnetic simulation results, demonstrating the accuracy of the MLP. Then, the genetic algorithm was combined with the MLP to search for the structures with $Ia\bar{3}d$ symmetry with a wide photonic gap width. For the titania material with dielectric contrast of 6.25, a maximum photonic gap width of 11.10% at a $f$ range of 0.9288-1.0380 $c/a$ was obtained. The performance was also calculated by MPB with spatial resolution of 32×32×32 and the structural parameters are shown in Table S13, which also pointed to the similar network with *lcs* topology (Figure 5H). Figure S28-S30 show the inverse design of photonic crystal structures to achieve the required $f_{mg}$ (0.8 $c/a$), gap width (15%) and material dielectric contrast 13.00. Based on our workflow, it is highly convenient to set various performance targets based on the band structure data predicted by the MLP to optimize the structural parameters of photonic crystals.

## 3. Discussion

The close correlation between performance and structure places high demands on structural modeling to achieve maximum design flexibility with as few structural parameters as possible.[52] However, this is difficult to achieve with traditional modeling methods.[53] The inverse Fourier transform, by freely adjusting the number of structure factors, can cater to the diverse structural design needs and effortlessly achieve a balance between the two requirements. Herein, the effective tuning of the structures was achieved using only 5 low-index reflections with a certain data range. Adding more high-index reflections and expanding the data range can improve structural resolution and structural modulation in more detail in future research. Moreover, all 230 space group geometries (not limited to cubic lattices) can be introduced to the Fourier method and arbitrary structures can be easily established and applied to our workflow by combining different Fourier components. In addition, our method lends itself towards generating disordered structures within a computational supercell. In such a situation, the Fourier coefficients would be restricted to a thin spherical shell of radius much greater than the reciprocal lattice constant of the supercell[38,54,55] with otherwise unconstrained *hkl*. This corresponds to a triclinic space group with a generally cubic unit cell, but in the absence of point symmetries. Since band structure calculations are very expensive in this scenario, MLP-assisted design might prove a vital ingredient in identifying and understanding disordered photonic crystals with large isotropic bandgaps.

The freely controllable number of structural parameters in Fourier design makes it convenient to combine it with the most fundamental fully connected MLP and achieve satisfactory prediction accuracy and structural design requirements without the involvement of complex neural networks in structural data processing. Here, our workflow does not involve complex and profound neural network architecture adjustment and modification of the underlying code, demonstrating the generalizability of the deep learning method as a tool in cross-disciplinary field research. The deep learning method is able to predict results at high speed with satisfactory accuracy requirements, which is different from the accurate description of physical phenomena by traditional electromagnetic simulation.[24,25,50,52,56-58] Particularly, by serving as a proxy mode for mathematical models describing physical phenomena, the efficiency in photonic structural design is thus greatly improved, opening pathways to the design of functional materials beyond bandgap optimization. This includes topological photonic crystals,[59] where a certain symmetry can predict band degeneracies with unique physical properties, but optimization is needed to achieve frequency isolation of these degeneracies,[60] and nonlinear photonic crystals,[61] for which phase matching is crucial. More generally, 3D metamaterial design[62] would greatly benefit from a slightly modified approach, replacing MPB with an eigensolver that can model the associated physical system.

Herein, our method has produced new discoveries in the study of photonic crystals. It was generally believed that face-centered cubic lattices with most spherical Brillouin zones would be favorable for the alignment of band gaps in all propagation directions to form a CPB[3,31,39,40] Besides, bicontinuous interwoven domains of high and low refractive index allow the electric field to concentrate best in the high index region, thereby maximizing the frequency difference between valence and conduction band.[27] For this reason, the single network topologies exhibit superior bandgap performance, while the corresponding double networks do not exhibit a CPB.[40] Our results also confirmed the excellent bandgap properties of SD-based structures with a minimum requirement of dielectric contrast and yielded the largest gap width.[39] Similarly, the SG-based structure (*srs* net) with symmetry $I4_132$ also showed excellent photonic bandgap properties. Notably, the *srs* structure is intrinsically chiral and features chiro-optical properties.[63] This aspect was not addressed in this article, but strong circular dichroism or optical rotation found in multi-*srs* topologies[47,64] could be improved with our approach. However, our study demonstrated that the 3D photonic structure with *lcs* topology with space group $Ia\overline{3}d$ deserves more attention. $Ia\overline{3}d$ is the space group of the DG structure (*srs-c* net) with a body-centered cubic lattice and consists of tetravalent nodes and hexagonal shortest rings such as the SD (*dia* net).[65] Remarkably, the *lcs* net is the skeletal graph of the so-called G' structure, a member of the $C(I_2-Y^{**})$ family discovered by von Schnering and Nesper.[20,65,66] Although a large bandgap in the $C(I_2-Y^{**})$ photonic crystal structure has been identified two decades ago,[45] this structure does not seem to have attracted much attention, and its network topology has not been revealed.

To further compare the photonic bandgap characteristics of the structures under these three space groups, the structures were further optimized under the constraints of dielectric contrasts of 6.25 and 13.00, respectively. As shown in Table 1, the gap width and gap frequency range of the *lcs* net are better than those of *srs*-based structures and even comparable with the *dia* net. While Bragg scattering comes from perfectly periodic structures, interference can also occur between scattering centers in the absence of crystallographic symmetry.[55] Potential candidates for CPB materials should be composed of geometrically identical scattering units because increasing the spatial similarity of the scattering units maximize the overlap of their spectral ranges that suppress propagation, thus facilitating the formation of CPBs. In 3D space, tetravalent and trivalent networks with low coordination number and maximum local self-uniformity have been demonstrated generate large CPBs.[38] From the structural point of view, the *lcs* topology contains all the factors needed to design an ideal photonic crystal. The lattice shows strong isotropy of the tetravalent scattering units with good, albeit not perfect, local self-uniformity. The exceptional bandgap performance of the *lcs* topology, particularly its minimum requirement for dielectric contrast with open CPBs comparable to SD structures (*dia* net), confers its unique advantages in the fabrication of optically controlled devices.

**Table 1.** Bandgap properties of photonic crystal structures with the largest photonic gap width in the structures with *dia*, *srs* and *lcs* topology optimized under dielectric contrast of 6.25 and 13.00, respectively. The corresponding geometries and band structures are shown in Figure 5H, S31 and S32, and were calculated by the MPB with the spatial resolution of 32×32×32.

| Topology | Dielectric contrast | $f_{mg}$ ($c/a$) | Gap width | Gap $f$ range ($c/a$) |
|---|---|---|---|---|
| *dia* | 6.25 | 0.6602 | 13.86% | 0.0915 |
|  | 13.00 | 0.6041 | 31.24% | 0.1887 |
| *srs* | 6.25 | 0.5651 | 8.07% | 0.0456 |
|  | 13.00 | 0.4867 | 27.47% | 0.1337 |
| *lcs* | 6.25 | 0.9834 | 11.10% | 0.1092 |
|  | 13.00 | 0.8283 | 24.37% | 0.2019 |

In the design process for structures with defined bandgap performance like $f_{mg}$ and gap width, multiple candidates are generated (Figure S15, S22 and S29), due to the fact that the similar $f_{mg}$ and gap width can be produced by many different structures.[67] This inherent one-to-many relationship can lead to conflicting training examples, thereby complicating the convergence of the training process.[57,67] In our workflow, the neural network was trained as a proxy model for electromagnetic simulation used to predict the photonic bandgap performance of the structure, effectively avoiding this problem. And by iterating the structural parameters through a multi-objective genetic optimization algorithm, the set of possible optimal solutions can be obtained when considering two objectives ($f_{mg}$ and gap width), which is known as the Pareto front.[26] It is well known that errors in the prediction results of neural network models are unavoidable. Additional performance validation and selection from the obtained set of candidates can help to further increase the probability of obtaining a final structure that meets the requirements. All these strategies ensure the effectiveness of our workflow in the design of photonic structures with predefined properties.

In our calculations, there are also a large number of hybrid structures with disconnected nodes coexisting with network structures showing large CPBs (Figure S15, S22 and S29). Based on the inspiration brought by the effect of connectivity on the band gap in 2D structures,[15,68] a variety of 3D photonic crystal-based methods have also demonstrated that not only network structures[37,69] but also disconnected geometries[32,70,71] are capable of forming CPBs. The hybrid structures that emerge in our work provide an extra complement to photonic structural design. At the same time, three-component structures are often experimentally obtained through backfilling a mold with high-index materials[72] but are rarely studied. The Fourier method is ideally suited to study such multi-component materials by introducing an additional threshold in Equation 1. Even more so, the Fourier sum describes a thresholdless continuous variation of permittivity that is computationally favourable as it allows the plane wave expansion in MPB to converge exponentially[73] and might lead to unprecedentedly large PBGs as it widens the parameter space. Moreover, our results demonstrate that large bandgaps are also robust to discrete substructure units in network structures. This complements the previous view that large bandgaps are inherently robust to the roughness of structural surfaces.[37,74] The above points indicate that our method reveals a rich diversity in photonic crystal architectures.

## 4. Conclusions

In summary, the integration of Fourier analysis with deep learning presents a novel methodology for the robust and adaptable modeling of periodic structures. This approach not only reveals a rich diversity in structural design, but also establishes a generalizable framework for exploring material properties across various disciplines. A compelling demonstration of this framework in the application of photonic crystal design successfully identifies substantial photonic band gaps within 3D photonic architectures and uncovers previously overlooked topologies such as *lcs*. The methodology represents a significant advancement beyond traditional design paradigms by enabling comprehensive exploration of all 230 space group geometries and facilitating the creation that extend beyond conventional knowledge boundaries. We envision this framework will catalyze groundbreaking innovations in structural engineering while offering a transformative toolset for advancing material science research and developing next-generation functional materials.

## Methods

*Structural modeling by inverse Fourier transform*

The photonic structures were modeled by inverse Fourier transform using structure factors $F_{hkl}$ of selected *hkl* reflections with reference to the calculation formulas in *International Tables for X-Ray Crystallography* (Table S1-S7).[19] Taking space group $Fd\bar{3}m$ as an example, structure factor amplitudes of 111, 220, 311, 222, 400 reflections and the isosurface threshold *t* with random values were used as structural parameters. Before inputting the equation, each unique reflection was expanded to all equivalent reflections and the crystal factor phases were changed according to crystal symmetry. Then the $\rho_{(r)}$ for all coordinates (*r*) in space were computed to form the level set. After normalization based on the minimum and maximum values of the electron density, the space with normalized electron density $\rho_{norm\,(r)}$ can be divided into two regions using a threshold *t*. In this study, it was defined that the material with dielectric contrast greater than 1 was occupying the space where $\rho_{norm\,(r)} > t$, and vice versa. Different geometries were obtained by changing space group symmetries and adjusting the geometric structural parameters.

*Photonic band structure calculation and dataset construction*

The computation of the photonic band structure was performed using MPB based on plane wave eigenmode search.[35] The *k*-paths in reciprocal space referred to the suggestions provided by Stefano Curtarolo.[48] And the resolution was set to 16. For the construction of the dataset, the first structure factor amplitude in the geometrical structural parameters was specified as 1, and the remaining structure factor amplitudes were assigned in the range of -1 to 1 with random values. The threshold *t* was randomly generated between 0 and 1, and the dielectric contrast was randomly assigned between 1 and 16. The geometrical structural parameters and dielectric contrast were inputted into the MPB and the corresponding photonic band structures as well as gap properties were calculated. The results of all calculations were summarized into the dataset.

*Neural network architecture and training*

The dataset consisting of 100 000 samples was randomly divided into a training dataset (70%), a validation dataset (20%) and a test dataset (10%). The open source PyTorch was employed to construct MLP with multiple hidden layers. The neural network model contained six neurons in the input layer, each corresponding to the amplitudes of the structure factor for the last four selected reflections, the threshold *t* and the dielectric contrast of the material. All input parameters were normalized to [0, 1] in their respective value ranges before feeding them to the neural network. The one-dimensional array output from the neural network output layer is reshaped into a two-dimensional array that corresponds one-to-one with the discrete frequency data of the photonic band structure. ReLUs were added as activation functions. Dropout was added after each hidden layer to prevent overfitting for neural network training. The loss was evaluated using MSE loss function

$$MSE = \frac{1}{n}\sum_{i=1}^{n}(y_i - \hat{y}_i)^2 \qquad (3)$$

where $n$ is the number of samples and $\hat{y}_i$ and $y_i$ are the predicted and true values for sample numbered $i$. Adam optimizer was used for training the neural network. The trained MLP was able to directly predict the corresponding photonic band structure data from the structural parameters, and the corresponding gap performance can be obtained through a simple post-processing step.

The hyperparameters (batch size, number of hidden layers, number of neurons in each hidden layer, learning rate, dropout size and number of epochs) were optimized using the Bayes method that comes with the wandb library. The optimized MLP contained 7 hidden layers and 2 048 neurons per hidden layer. 5 000 samples were fed to the neural network per batch during training. The learning rate of the Adam optimizer and the dropout size were set to 0.00001 and 0.1. After 3 000 epochs, the loss of MLP almost stopped decreasing.

*Inverse design of photonic crystal structures*

The NSGA-II[26] provided by the open source multi-objective optimization algorithm library pymoo[75] was used in combination with the trained MLP for inverse design of photonic crystal structures with predefined properties. The general approach was to search the structural design space of the photonic crystal, namely the amplitudes of the four structural factors mentioned above and the threshold $t$, using the predefined mid-gap frequency and gap width as the optimization objectives and the specified material dielectric contrast as the constraint. Iterations were performed to minimize the MSE between the gap performance predicted by MLP and the predefined optimization objective. The inverted generational distance (IGD) was employed as a metric to simultaneously evaluate the convergence and diversity of the optimization process. The inverse structural design processes were performed on two 3.2-GHz Intel(R) Xeon(R) 6146 CPUs and an NVIDIA Quadro P2000 GPU for 120 min. The obtained structural parameters were used to construct the structure by inverse Fourier transform and the final band structure was calculated using MPB at a spatial resolution of 16×16×16 for initial search or 32×32×32 for specific structures.

**Acknowledgements**

This work was financially supported by the National Natural Science Foundation of China (Grant No. 22425303, 22472058) and the Fundamental Research Funds for the Central Universities.

**Keywords**

structural design, Fourier analysis, deep learning, photonic crystal, complete bandgap